\documentclass[%
 reprint,
 amsmath,amssymb,
 aps,
floatfix,
]{revtex4-1}

\usepackage{graphicx}
\usepackage{dcolumn}
\usepackage{bm}

\begin{document}

\preprint{APS/123-QED}

\title{Propagation of Fluctuations in Au+Au Collisions at FAIR energy}

\author{S. Ahmad}
\email{s.bhat@gsi.de} 
\affiliation{University of Kashmir, Srinagar India}
\author{M. Farooq}
\affiliation{University of Kashmir, Srinagar India}
\author{H. Jahan}
\affiliation{Aligarh Mushlim University, Aligarh India}
\author{S. Chattopadhyay}
 \email{sub@vecc.gov.in}  
\affiliation{Variable Energy Cyclotron Centre, Kolkata India}
\author{S. Bashir}
\affiliation{University of Kashmir, Srinagar India}
\author{N. Ahmad}
\affiliation{Aligarh Mushlim University, Aligarh India}

\date{\today}

\begin{abstract}
Event by event fluctuations of particle multiplicities and their ratios are considered to be sensitive probes to the exotic phenomena in high energy heavy ion collisions like phase transtion or the occurence of critical point. These phenomena might take place at different time after the collision based on fulfilling the required conditions at a particular time. Fluctuations are therefore expected to show non-monotonic behaviour at the of time of occurence of these phenomena. Experimentally, fluctuations are measured at freezeout. In this work, using the hybrid version of the UrQMD event generator, we have investigated the propagation of fluctuations of particle multiplicities, their ratios and the ratio of total positive and negative charges in AuAu collisions at E$_{lab} < $ 90 AGeV. Two commonly used experimental measures i.e., $\sigma^2$/mean and $\nu_{dyn}$ have been used in the analysis in a given acceptance. The hybrid model, i.e., UrQMD with hydrodynamic evolution has been used to  study the effect of hydrodynamic evolution on these conventional fluctuation measures. It is observed that the fluctuations as measured by $\sigma^2$/mean and $\nu_{dyn}$ gets reduced considerably at freezeout. The dominat structures present at the initial stage of the evolution get smoothen out. However, the energy dependence of the fluctuations remain preserved till the freezeout. The hydrodynamic evolution of the model with chiral equation of state shows considerably higher fluctuation at lower collision energy as compared to pure hadronic transport version or the hybrid version with hadronic equation of state. The time evolution of the higher order moments of net-proton distributios for particles in a specified coverage showed similar behaviour.

\end{abstract}

\maketitle


\section{Introduction}
In high energy heavy ion collisions studied at RHIC and LHC, it has been concluded that partonic media have been created ~\cite{qgp-rhic,qgp-lhc}. As per Lattice simulations, the transition from hadronic to partonic phase is believed to be a cross-over. Various results indicate that the matters found at RHIC and LHC behave like a liquid with very low values of the ratio of shear viscosity to entropy suggesting this to be a perfect liquid~\cite{qgp-fluid}. Even though, presumably due to the continuous nature of the phase transition at RHIC and LHC, no 'smoking gun' signature of the phase transition has been observed, however, among the myriads of proposed observables of such phase transitions, one particular signature is believed to give unambiguous signals for interesting phenomena like the phase transition or the critical point is the event by event fluctuations of the conserved quantities. Experiments are being performed or being planned to explore the matter at higher net-baryon density and low temperature to search for the phase transition away from the cross-over which could take us towards the critical point, the end point of cross-over and begining of 1$^{st}$ order phase transition.

In these collisions, the statistical models explain the global properties of particle production reasonably well. In the context of such a picture, the fluctuations of conserved quantities, which could be connected to the susceptibility, meaning the response to external influence, are sensitive to the phenomena like the phase transition or the critical point \cite{Koch}. At T$_c$, the phase transition temperature, derivatives of free energy diverge inducing fluctuations to show clear discontinuity. For 2$^{nd}$ order phase transition, for example, the fluctuation reduces drastically if matter freezes out at T$_c$ \cite{pt2}. On the other hand, for a first order phase transition, fluctuation increases sharply due to droplet formation \cite{pt3}. This variable has been said to  thus classify the order of phase transition.  

Experimentally, the variable of interest is measured event by event and the moments of the distribution of the variable concerned provide the measure of its fluctuations. The quantity of interest in the context of fluctuation measurement depends on its sensitivity to the phenomena of interest. The range of quantities for which fluctuations have been measured include multiplicities of produced particles, event-wise ratios of yields of particles, total measured energy (or E$_T$) in an event, particle-averaged transverse momentum in an event among others. It has been discussed that for conserved quantities,  for good stable study, the detector acceptance plays a major role in the measurement of fluctuations. Over full acceptance, conserved quantities do not show any fluctuation, the detector coverage should be reasonably large to accept a large fraction of particles of interest produced in the event. 

Measurments of various observables have been performed at SPS, RHIC and LHC  describing different kind of correlations. Among them, prominent ones include $\Phi_q$ of NA49 \cite{tp3}, $\nu_{{\pm},dyn}$ of STAR \cite{tp4} and $\nu$(Q) as well as $\nu_{{\pm},dyn}$ used by PHENIX \cite{tp5}. A common framework has been made to relate all these variables \cite{tp6,tp7}. Based mainly on the fluctuations measured using the 2$^{nd}$ moment of the distributions, and studied with respect to centrality and $\sqrt{s}$ of the collisions,  no result so far has shown a large discontinuity in the net-charge fluctuations.

One common complexity associated with the fluctuation measurement in high energy heavy ion collisions is to construct proper variable which is free from effects like efficiency, acceptance, impact parameter fluctuations among others. Some of the observables being used extensively in analysing the data have been discussed in the next section. One approach to eliminate the impact parameter fluctuation, commonly known as volume fluctuation is to take the ratio of event by event yields of two types of particles e.g., K/$\pi$, p/$\pi$. Fluctuation of K/$\pi$ ratio and their charge dependence is of particular interest as they represent the fluctuations of strangeness production. The $\sqrt{s}$ dependence of K/$\pi$ ratio at midrapidity measured at AGS and SPS energies shows a "horn"-like structure at $\sqrt{s} \approx $ 7 GeV~\cite{tp81}. This interesting structure has been explained by invoking a QGP-based model that enhances the strangeness significantly at the point of transtion. Measurements have been performed on the event by event fluctuation of K/$\pi$ ratio  and their charge dependence at SPS, RHIC top energy and RHIC beam energy scan (BES). While SPS shows sharp fall of fluctuation with $\sqrt{s}$ before saturation~\cite{tp81,tp82}, the RHIC BES does not show any such rise at similar $\sqrt{s}$~\cite{tp9}.

It should be mentioned that the connections of the physical quantities like phase transition, critical point with measured fluctuations are relevant for a thermalized system, in which fluctuation of conserved quantities(Q) is related to susceptibility via $<\delta Q_i \delta Q_j>$ = VT $\chi_{i,j}$. Some of the quantities being measured experimentally have been extracted in the statistical simulation of QCD in lattice.
Results of statistical simulation of QCD on lattice showed that higher order moments like skewness and kurtosis are having better sensitivity to the critical point due to higher power dependence on correlation length ($\xi$) \cite{sk1,sk2}. The dependence of skewness (S) goes as $\xi^{4.5}$ and for kurtosis(k) it goes as $\xi^{7}$. In the measurements done at RHIC, variables like $\kappa$$\sigma^2$ and S$\sigma$ have been derived for net-proton and net-charge measured at different $\sqrt{s_{NN}}$. With these variables, data could be compared directly with lattice results\cite{sk3}.

Experimentally, the fluctuations of quantities of interest are measured at freeze-out. Depending on the time of origin of the interesting phenomena leading to enhancement or reduction of fluctuations, it may propagate over the period of particle production, equilibration and space-time expansion. 
If for example, a transition takes place at the pre-equilibrium stage that leads to a large fluctuation of the yield of conserved quantities at that moment, the system will then evolve typically through 6-10 fm/c in case of RHIC/LHC and similarly at lower energy and then the fluctations might get diluted. Recently \cite{shk2}, investigations have been made to study the propagation of a point like disturbance initiated at 1 fm/c till the freeze-out. Detailed hydrodynamical calculation shows that the disturbance results in an azimuthal shape which could be explained to be a convolution of various flow components e.g. v$_2$, v$_3$, v$_4$. 
The equilibration and the time of its occurence is an important point of study both experimentally and theoretically in the field of high energy heavy-ion collisions. The equilibration time depends on the beam energy and varies from below 1 fm/c at LHC to a large value at SPS. In most of the works on the propagation of fluctuation, through the medium, the fluctuations have been made to traverse through an equilibrated medium. One source of fluctuation before the equilibration is the initial geometry fluctuations or the fluctuation of the gluon density of the colliding nuclei. 

For heavy-ion collisions at relatively lower energy, likely to be accessible at FAIR, as per the transport model calculations, a baryon density of the order of 1.5/fm$^3$ is likely to be produced for a very short period during the overlap of the nuclei before the system reaches equilibration. At such a high density, the collision zone might transform into a partonic medium which will eventually equilibrate and then convert into hadrons. The extreme density region might result in large fluctuations of specific quantities as the density also fluctuates event by event. It will be useful to study the propagation of these fluctuations in terms of their survival at freezeout.

In this work, we have used UrQMD 3.3 version to study the space-time evolution of the fluctuations in Au-Au collisions at  E$_{lab}$ ranging from 10 AGeV to 90 AGeV. Two modes of UrQMD have been used, one with Hydro evolution (known as hybrid model) and another without the inclusion of hydro expansion. In this work, we have looked into the event by event fluctuations of the particle yields or the fluctuations of the correlated production of particles by measuring the event by event ratio of particle yields. This work does not include the studies of moments of the azimuthal distributions as a measure of fluctuations. 

The document is organised as follows, in section-II, we give a brief introduction to UrQMD. We discuss the methodology of this analysis and the descriptions of the variables used in this analysis in section-III. In the next section, we present results of the time evolution of fluctuations of kaon and pion multiplicities, of their ratio and of the ratio of multiplicities of positively and negatively charged particles (Q$^+$/Q$^-$). We have also extracted the results for those fluctuations at freezeout and in presence of hydrodynamical expansion. In section-V, we have discussed the time evolution of teh net-proton distributions starting from the mean value to the higher order moments. The work has been summarized in section-V.

\section{Ultra Relativistic Quantum Molecular Dynamics(UrQMD)}

UrQMD is one of the models being used widely in describing the results of high energy heavy-ion collisions. The initial version of the model was based on the use of transportation of hadrons with the implementation of various intermediate phenomena in the model. The model includes transportation of various degrees of freedom (e.g. baryons and mesons) and the production of new particles and their interaction. It treats the production of paricles via fragmentation of strings made of valence quarks of the original colliding hadrons \cite{urq1,urq2,urq3}.  The model did not have any equilibration implemented and therefore represents the non-equilibrated mode of transport. The final hadrons might be the decay product of hadronic resonances formed during the transportation. It should be noted that no phase transition could be implemented in this form of the model, which has found wide applications in explaining the data like particle multiplicity, their species dependence, energy flow, azimuthal distributions among others. Different properties of the medium like net-baryon density could be calculated during the transportation. As the model implements only the hadronic form of the medium, so it has been used at several places to represent the hadronic reference of different observables. The model has been used to explain data from a wide range of collision energy i.e. from SPS to RHIC.

The energy and species of the colliding particles are given as the input to run the UrQMD code and the output could be finally detectable particles or the 'collision' file containing the time history of the collision. The parameters like the time of freeze-out could be given as input.

In the 'hybrid' version of UrQMD, the transportation of hadronic medium is stopped and a hydrodynamical flow is set in \cite{urq4,urq5,urq6}. The equation of state represents the type of medium under the hydrodynamical expansion. The time after which hydrodynamical flow sets it could be set as an input parameter but is set as default to be the time of crossing of two Lorentz-contracted nuclei via,  

\begin{equation*}
t_{start} = \frac{2R}{\sqrt{\gamma^{2}-1}}
\end{equation*}

After the initial pre-equilibrium stage of UrQMD with the initial state of the collision implemented, the subsequent hydrodynamic evolution is performed using the SHASTA algorithm \cite{urq7,urq8}.  In this algorithm, cells are formed with specific thermodynamic parameters to implement the hydro transport. There are provision of using two types of equation of states i.e., hadronic EOS which does not have any phase transition and the chiral EOS with the phase transition built in. In the 'hydro' model, the hydrodynamic evolution is stopped after the energy density reaches below some given threshold as compared to the ground state energy density (e.g. $\sim 730 MeV/fm^{3}$). Finally particles are formed by using the Cooper-Frye equation on an isochronous hyper-surface of the fields under hydrodynamic evolution. The hadrons thus formed might interact again using the hadron cascade before being detected. It has been shown the model could be used over a large range of collision energy i.e. E$_{lab}$ of 1 GeV to 160 AGeV at SPS and then at RHIC \cite{urq10}.

\section{Methodology}

The transport model UrQMD has been used to produce particles at different time slices after collision. The particle yields at different time bins are analyzed to study the time evolution of the event by event fluctuations using the measures discussed below. In this analysis, particles in the pseudorapidity region of $\pm$ 1 around midrapidity are taken for further analysis.

 The present work is focussed on nuclear collisions at FAIR energy range producing matters of high net-baryon density at the time of overlap of the nuclei. As the time progresses, net-baryon density decreases, it will therefore be useful to study if the fluctuations present at the time of collision remain unaltered with time. We have focussed on the density dependence of the fluctuations of particle multiplcities, their ratios and of net-charge. In general the procedure that is usually adopted in such a study is to introduce a fluctuation from outside to study its propagation with time. In this study however, we refrain from doing this, as the procedure of introduction of an external source of fluctuation is debatable and vary from one physical process to other. We have however studied the fluctation as it comes in an expanding medium as given by the transport model. 

In this study we have used two version of UrQMD, in the conventional UrQMD (hadronic version), time scans have been performed from t=1 fm/c till t=10 fm/c. It should be noted that there is no phase transition implemented in this version. The temporal variation of the fluctuation is due to the evolution of the hadronic medium as created by the transport model. In this model, no explicit implementations have been made for chemical or thermal equilibria. In another version of UrQMD (hybrid version), as discussed in the earlier section, transport of particles are stopped and hydrodynamical evolution is performed with a given equation of state. In the version of UrQMD used here, 3+1 dimensional evolution has been implemented for hydrodynamics. Both the hadronic and chiral equation of states have been used. The freezeout has been represented by particle production using Cooper-Frye formalism.

In this work, following studies have been performed: (a) the evolution of fluctuations represented by the measures discussed below have been studied at time bins of 1 fm/c width from the time of collision upto 10 fm/c. (b) the fluctuations of a set of variables are evaluated before the initiation of hydro and then compared with the results from the same variables after hydro evolution (c) the effect of the change of equation of state has been studied by comparing the fluctuations of the quantities concerned at freezeout for three cases, pure transport model without the use of hydrodynamical evolution and then with hydro evolution using two equation of states. 

We discuss below the construction of a few variables used as measures of fluctuations in this analysis.
The aim of the construction of specialized variables is to eliminate fluctuations from trivial effects like statistical fluctuation, volume fluctuations among others. Depending on the type of the observable being studied, the effects might be different. In this work, two sets of quantities have been studied, the multiplicities themselves and their ratios. In the present work, for multiplicity fluctuation study, we have used the variable $\sigma^2$/mean, which is expected to be unity for independent particle emission following poissonian statistics. For ratio fluctuation study, we have used a variable known as $\nu_{dyn}$, that is being used exensively in experimental data analysis and is said to a robust observable for the study of ratio fluctuations.
 $\nu_{dyn}$ was originally introduced to study net charge fluctuations \cite{vd1, tp4}. The $\nu_{dyn}$ in its final form is given by equation below, where m and n represent multiplicities of two species.

{\footnotesize
\begin{equation}\label{vd}
\nu_{dyn} = \frac{<m(m-1)>}{<m>^{2}} + \frac{<n(n-1)>}{<n>^{2}} - 2 \frac{<mn>}{<m><n>}
\end{equation}
}

To describe the dynamical fluctuation in $K/\pi$ ratio, we can write
{\scriptsize
\begin{equation*}
\nu_{dyn, K/\pi} = \frac{<N_k(N_k-1)>}{<N_k>^{2}} + \frac{<N_{\pi}(N_{\pi}-1)>}{<N_{\pi}>^{2}} - 2 \frac{<N_{K}N_{\pi}>}{<N_{K}><N_{\pi}>}
\end{equation*}
}

where $N_{K}$ and $N_{\pi}$ are the event-wise number of kaons and pions in a given acceptance, respectively. Similar formula can be constructed for Q$^+$/Q$^-$, $p/\pi$, $K/p$ and other ratios. By definition, $\nu_{dyn}$=0 for the case of a Poisson distribution of kaons and pions. It is also largely independent of detector acceptance and effeciency in the region of phase space being considered \cite{tp4, vd2}.

In earlier experiments another measure called $\sigma_{dyn}$ \cite{vd3} has been used for particle ratio fluctuation. $\sigma_{dyn}$ is defined as, 
\begin{equation*}
\sigma_{dyn} = sgn (\sigma_{data}^{2} - \sigma_{mixed}^{2})\sqrt{|\sigma_{data}^{2} - \sigma_{mixed}^{2}|}
\end{equation*}
where $\sigma$ is the width of the ratio distribution in either data or in mixed events. Two variables have been found to be related~\cite{vd6} as
\begin{equation*}
\sigma_{dyn}^{2} \approx \nu_{dyn}
\end{equation*}
In this study, we have used $\nu_{dyn}$ as a parameter to study the fluctuations in different ratios like $K/\pi$ and $Q_{+}/Q_{-}$.

\section{Results}

\subsection{Time evolution of net-baryon density, particle multiplciity and fluctuations}

\begin{figure}[htbp]
  \centering
  \includegraphics[width=0.45\textwidth]{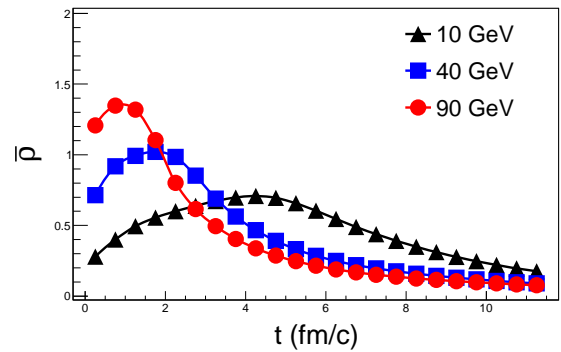} 
  \caption{Evolution of baryon density with time elapsed after collision}
  \label{density-with-time}
  \end{figure}

Fig.~\ref{density-with-time} shows the time evolution of the net-baryon density at three different incident energies. The flat top varies with the incident energy being longer-lived at lower energy. The density reaches a value upto 10 times the nuclear matter density suitable for a transtion to a non-hadronic state. Fig.~\ref{density-with-time}, which shows the variation of net baryon density with time at different collision energies, demonstrates the importance of heavy-ion collisions at FAIR energies that generate high density matter. As expected from the time required for two nuclei to pass each other, the peak density is achieved faster at higher beam energy. It will therefore be very interesting to study the variation of particle multiplicity with time and its correlation with the net-baryon density. It should be mentioned that, the time of reaching the peak net-baryon density might not coincide with the equilibration time especially at lower collision energy. A high density matter undergoing a transition to a partonic state at the highest net-baryon density might undergo rescattering to reach equilibration at a later time. The signal like fluctuation of specific quantities might then undergo evolution to reach the freezeout stage via equilibration. The main aim of this work is to study this effect.

\begin{figure}[htbp]
  \centering
  \includegraphics[width=0.45\textwidth]{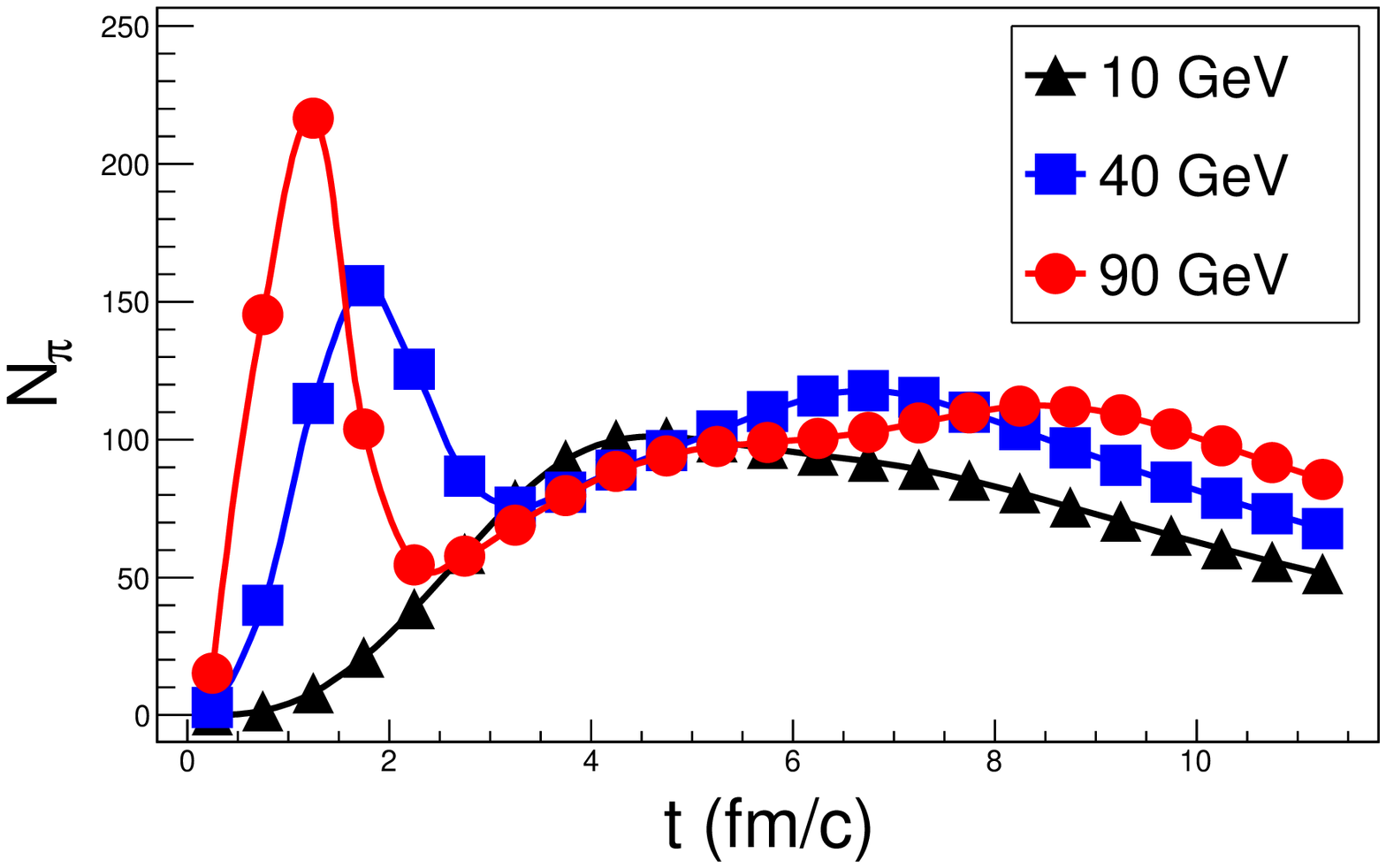} 
 \includegraphics[width=0.45\textwidth]{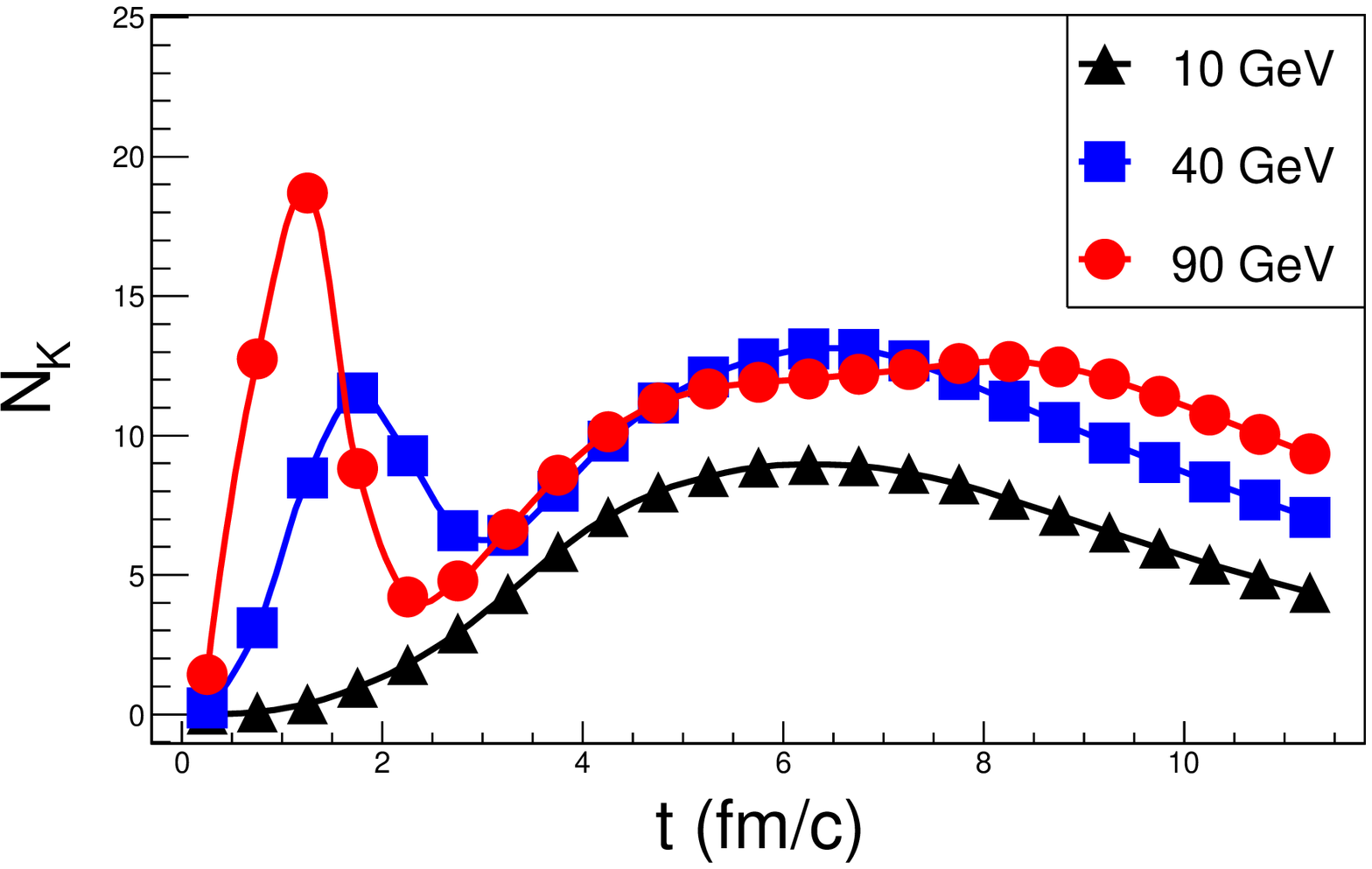} 
  \caption{mean multiplicities of pion and kaon with time}
  \label{mult-kpi}
  \end{figure}

\begin{figure}[h]
  \includegraphics[width=0.45\textwidth]{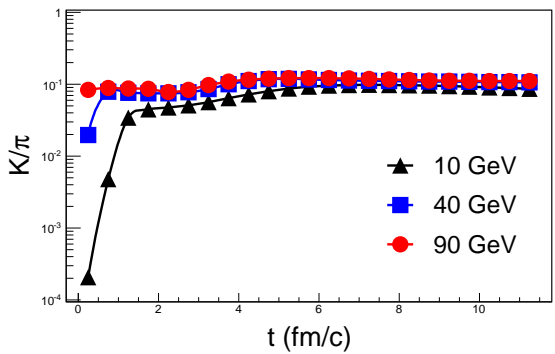} 
  \caption{Evolution of $K/\pi$ ratio with time}
  \label{ktopi}
  \end{figure}

Fig.~\ref{mult-kpi} shows the variation of  average multiplicities with time for pions, kaons at different beam energies. It is clearly evident from the figure that the multiplicities reach maxima at highest net-baryon density, the time of complete overlap of two nuclei. It subsequently reduces and then reaches stable values. This is more prominent at higher collsion energy, when the net-baryon density changes relatively quickly with time. At 10 AGeV, the multiplicity increases slowly to reach near satuartion values without the presence of a sudden variation with time. At that energy, the density also shows a smooth behaviour. Even though the particle production goes through peaks and valleys, the ratio of kaon and pion multiplicities remain almost constant throughout the progress of the medium as seen in Fig~\ref{ktopi}.

At lower incident energies, as the time of overlap is longer, the multiplicities reach respective peaks at later times, however, the subsequent reduction in multiplicities is slower compared to the case at higher energy collisions. A comparison between Fig.~\ref{mult-kpi} and Fig.~\ref{ktopi} shows that while the fluctuations in subsequent time bins of individual particle multiplicities are stronger for a medium in progress, the ratio of kaon to pion yield however shows weaker bin to bin fluctuation. It is therefore important to quantify the fluctuation using the widely used variables like $\sigma^2$/mean for particle multiplicities and $\nu_{dyn}$ for particle yield ratio.

 \begin{figure}[h]
  \includegraphics[width=0.45\textwidth]{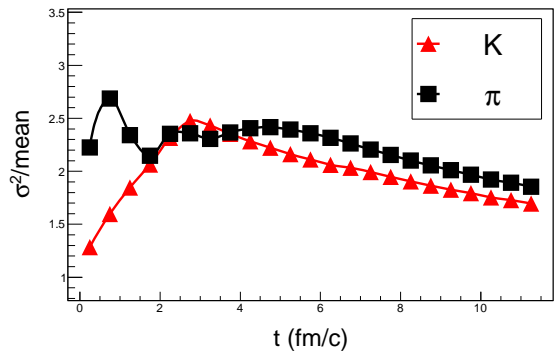} 
  \caption{$\sigma^2$/mean with time for $K$ and $\pi$ at 40AGeV Au+Au collisions}
  \label{sigma2-mult}
  \end{figure}


Fig.~\ref{sigma2-mult} shows the variation of $\sigma^2$/mean for kaons and pions with time at E$_{lab}$ = 40 AGeV. 
The evolution process dilutes the bumpy nature of the distribution to arrive at a relatively smoother fluctuation measure. However, it is clear that the particle-wise ordering of the initial fluctuation at highest density is preserved even after freezeout. The increase in pion multiplicity fluctuations with energy as compared to that of kaons is most likely to be due to pion productions due to resonance decays.

\begin{figure}[h]
  \includegraphics[width=0.45\textwidth]{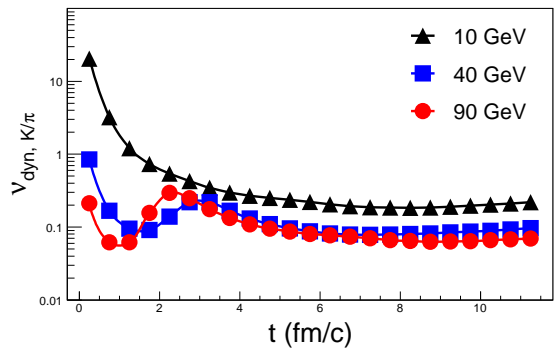} 
  \caption{Time evolution of $\nu_{dyn}$ of $K/\pi$ at different collision energies}
  \label{nudyn-kpi-energy}
  \end{figure}

\begin{figure}[h]
  \includegraphics[width=0.45\textwidth]{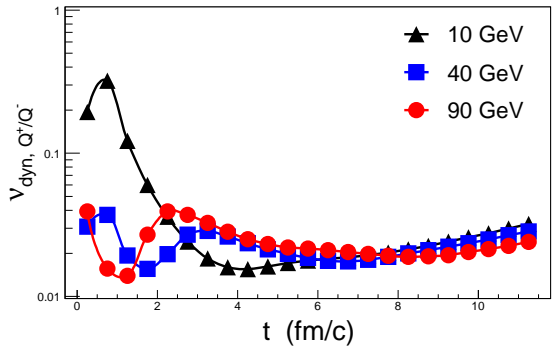} 
  \caption{Time evolution of $\nu_{dyn}$ of Q$^+$/Q$^-$ at different collision energies}
  \label{nudyn-qpqn-energy}
  \end{figure}

The time dependence of $\nu_{dyn}$ are shown in Fig.~\ref{nudyn-kpi-energy}~and~\ref{nudyn-qpqn-energy} for Au-Au collisions at three different energies for K/$\pi$ and Q$^+$/Q$^-$ respectively. $\nu_{dyn}$ is positive in both the cases showing effects of correlation term being smaller compared to the fluctuation terms in Eqn.(1). In case of K/$\pi$ , fluctuation is more at lower energy mostly due to long-lived dense phase and production of strange particles near threshold. This property could be used to study the euqation of state of the matter using the net-baryon density. 

In both the cases, we see interesting structures. $\nu_{dyn}$ reduces to a minima at the time of highest net-baryon density and then increases to reach near-saturation values before with a peak at intermediate time bins. Initial decreasing trend and dip at highest net-baryon density suggests a correlated production of kaons and pions which is dominant at peak net baryon density. With time, kaons and pions are produced independently thereby increasing $\nu_{dyn}$. Interestingly, the initial dip does not persist at a later time after collision.
It is seen from Fig.~\ref{nudyn-kpi-energy} that with increasing beam energy, the dip gets more prominent and occurs at earlier times. At very low energy, the peak and dip structure disappears and we see a smooth decreasing trend at E$_{lab}$=10 AGeV. The energy dependent ordering is however preserved till freezeout more prominently for K/$\pi$ even though there are some cross-overs during transportation. The absolute values of fluctuations as measured by $\nu_{dyn}$ for both the cases increase towards lower beam energy.  As the $\nu_{dyn}$ for K/$\pi$ essentially represents the fluctuation in strangeness production,  higher fluctuation at lower beam energy represents the fact that the multiple interaction, the prominent mechanism of production of strangeness at lower energy also leads to large fluctuations. For the case of Q$^+$/Q$^-$ initial charge conservation might have produced the correlated production of oppositely charged particles and the correlation reduces with resonance decays and of spreading of particles to higher rapidity range. 

\section{Multiplicities, ratios and their fluctuations at freezeout}

\begin{figure}[h]
  \includegraphics[width=0.45\textwidth]{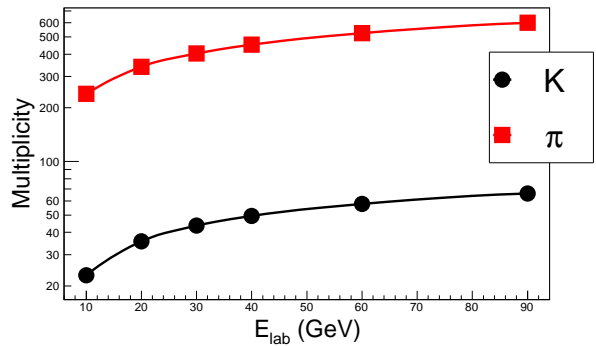} 
 \includegraphics[width=0.45\textwidth]{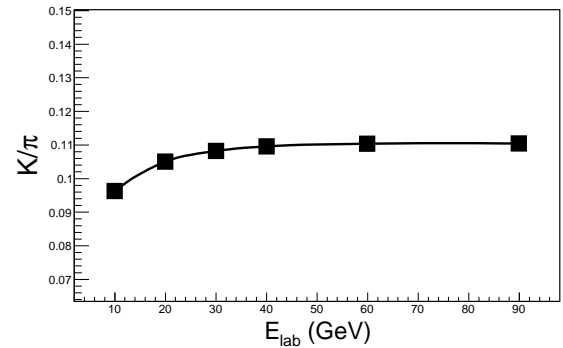}
  \caption{Beam energy dependence of kaon and pion multiplicities and their ratios at freezeout}
  \label{mult-elab}
  \end{figure}

As the time progresses, the multiplicities reach near-saturation values continuing till freezeout. The energy dependence of multiplicities at freezeout is shown in Fig.~\ref{mult-elab}.

The ratio of strange and light mesons plays an important role in understanding the medium created in high energy heavy-ion collisions. This ratio provides information on strangeness enhancement. The fluctuation of this ratio has been studied at SPS and RHIC energies. The fluctuation measured using $\sigma^2_{dyn}$ at SPS shows a decreasing trend with beam energy saturating at top SPS energy continuing to top RHIC energy.

 \begin{figure}[h]
  \includegraphics[width=0.45\textwidth]{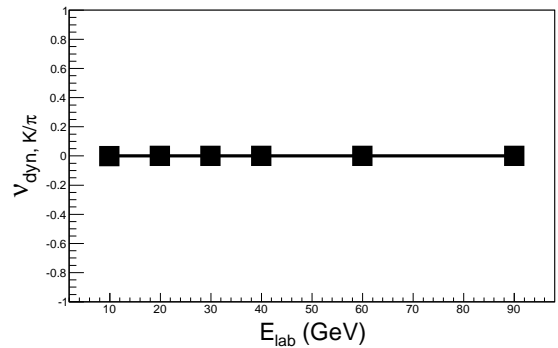} 
  \caption{$\nu_{dyn}$ of $K/\pi$ at freezeout}
  \label{nudyn-kpi-freeze}
 \end{figure}

Fig.~\ref{nudyn-kpi-freeze} shows the energy dependence of $\nu_{dyn}$ at freezeout for K/$\pi$ ratio. It is positive for K/$\pi$ showing the less-correlated production of  kaons and pions at a later time after collision.
We have seen that when we selected 1\% events of highest net-baryon density and compared $\nu_{dyn}$ for K/$\pi$ and Q$^+$/Q$^-$ with the bulk of events. They do not show significant differences in the time evolution of the fluctuations compared to the value averaged over all events.

\subsection{Effect of hydrodynamical evolution on fluctuation}

\begin{figure}[h]
  \includegraphics[width=0.45\textwidth]{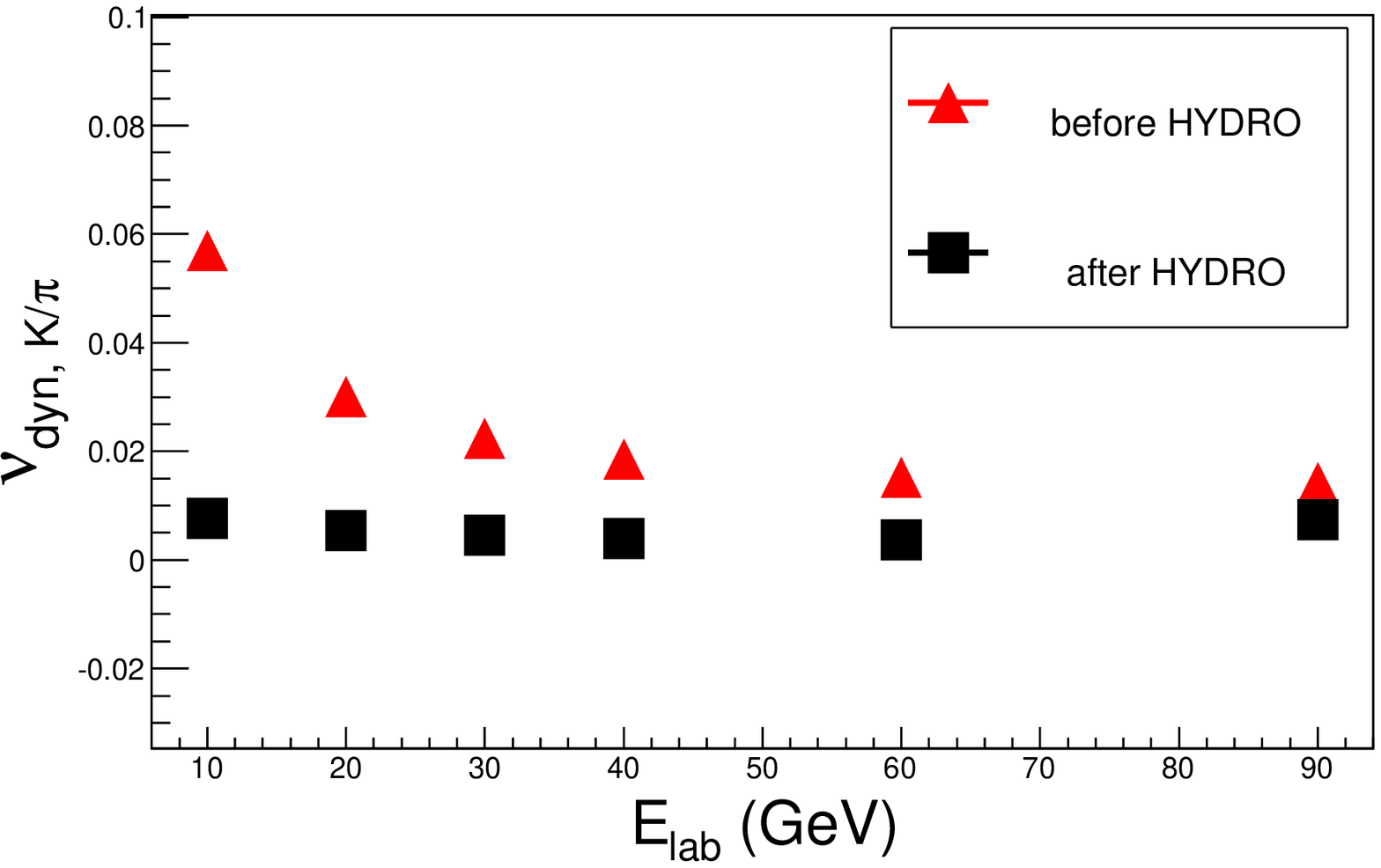} 
\includegraphics[width=0.45\textwidth]{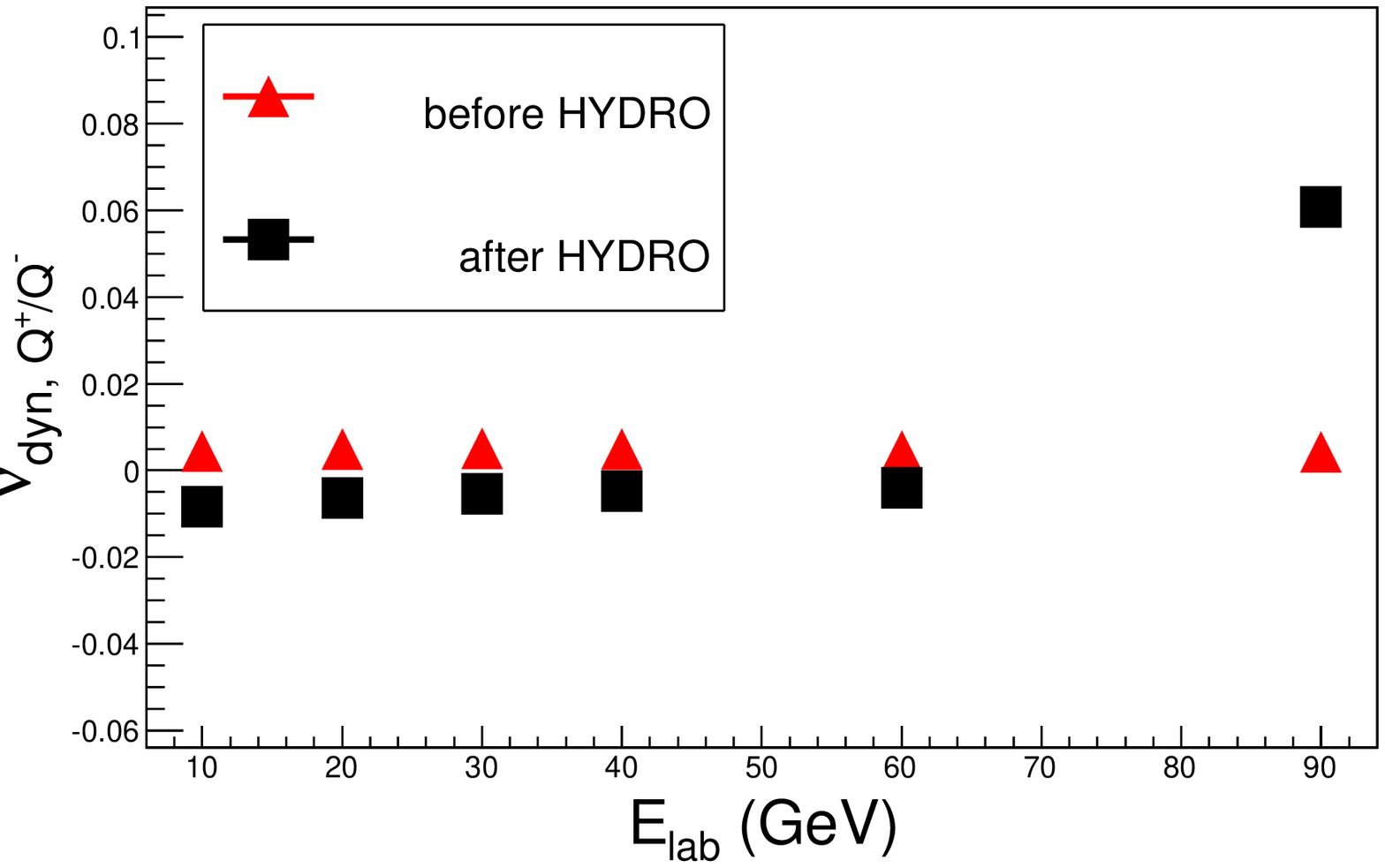} 
  \caption{Beam energy dependence of $\nu_{dyn}$ for $K/\pi$ and Q$^+$/Q$^-$ before and after hydro evolution}
  \label{nudyn-hydro-kpi}
  \end{figure}

Recently the role of hydrodynamic evolution to the initial fluctuation present primarily at the geometrical overlap of the colliding nuclei are being discussed extensively. In those discussions, inferences have been drawn on the signatures of azimuthal distributions of the produced hadrons that have been reflected by the higher harmonics of the azimuthal distributions.
In the present discussions, we study measures of the conventional meaures of fluctuations for particle yields and their ratios. The hydro evolution with hadronic equation of state has been switched on at time 5.56, 3.94, 3.22, 2.79, 2.28, 1.86 (fm/c) for beam energies 10, 20, 30, 40, 60, 90 (GeV) respectively, which continues till the lifetime of the hadronic phase. The conversion of fields to particles then take place using the Cooper Frye formalism after which rescattering processes implemented in UrQMD takes over. Fig.~\ref{nudyn-hydro-kpi} shows the energy dependence of $\nu_{dyn}$ for K/$\pi$ and Q$^+$/Q$^-$ before and after the hydro evolution. It is seen that for K/$\pi$, $\nu_{dyn}$ reduces considerably after hydro evolution more prominently at lower collision energy. This is presumably due to equilibration of the system for hydro evolution. The effect is stronger at lower energy as the duration of overlap is longer. The fluctuation in the ratio reduces with E$_{lab}$ in both cases (before and after hydro), however, the stronger energy dependence before hydro gets considerably diminished after the hydro evolution. At higher E$_{lab}$, where the fluctuation itself is considerably reduced, values from both the scenarios overlap.

In case of net-charge, even though the absolute values of $\nu_{dyn}$ are close to each other, however, after hydro evolution, it attains a negative value suggesting higher correlation between event by event yields of positively and negatively charged particles. It therefore sugests that with hydro evolution and subsequent production of particles by Cooper-Frye procedure, particles are produced with higher correlations between the oppisite charges. 
$\nu_{dyn}$ results at RHIC shows negative $\nu_{dyn}$ at all centralities. It is interesting to see that the final values at freezeout is negative as seen by experiments, however this has evolved from a positive $\nu_{dyn}$. It should be noted that for a case before hydro evolution, $\nu_{dyn}$ is positive throughout. It is however, necessary to check if the time-evolution without hydro reduces $\nu_{dyn}$ by the observed amount.

\begin{figure}[h]
  \includegraphics[width=0.45\textwidth]{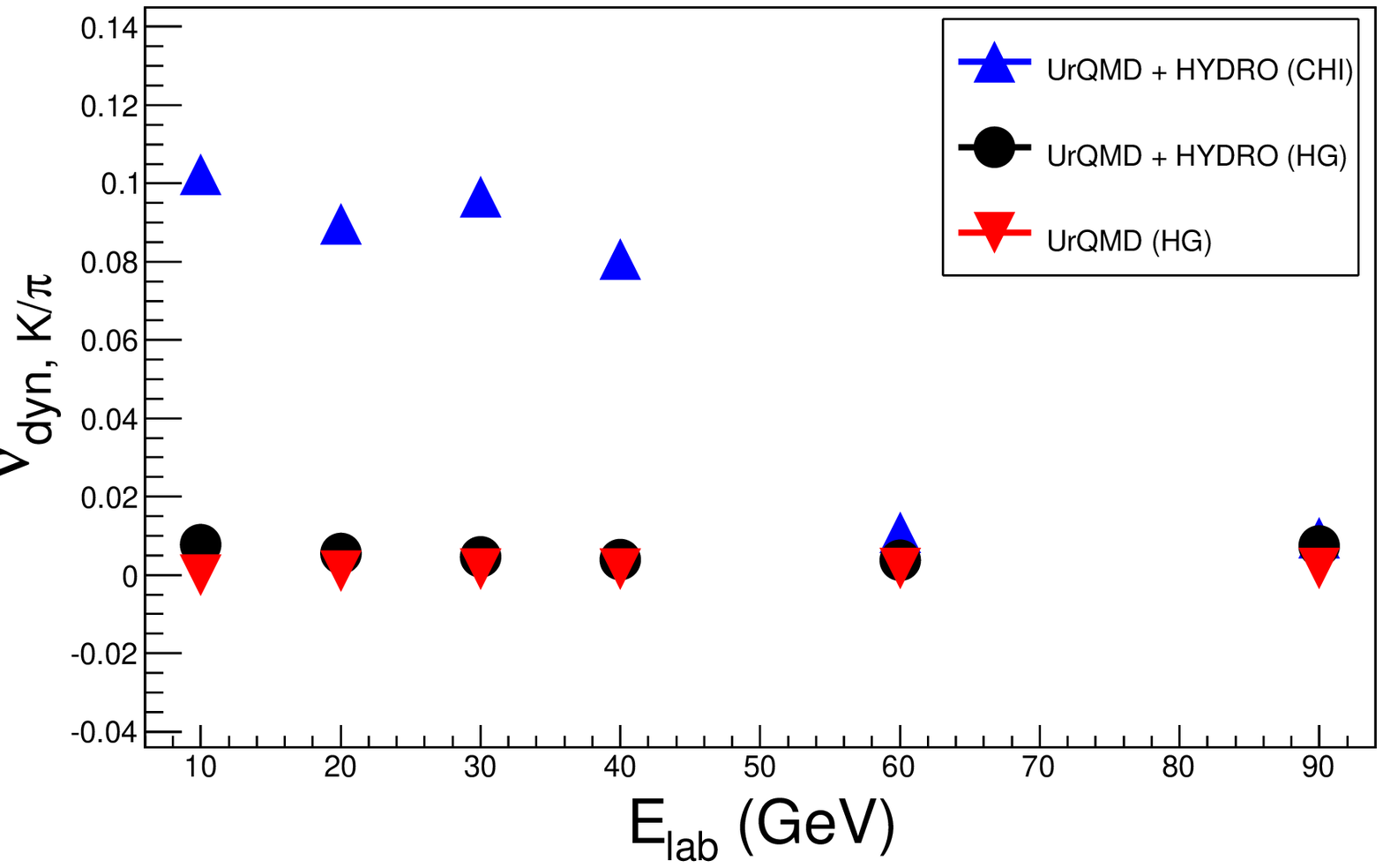} 
\includegraphics[width=0.45\textwidth]{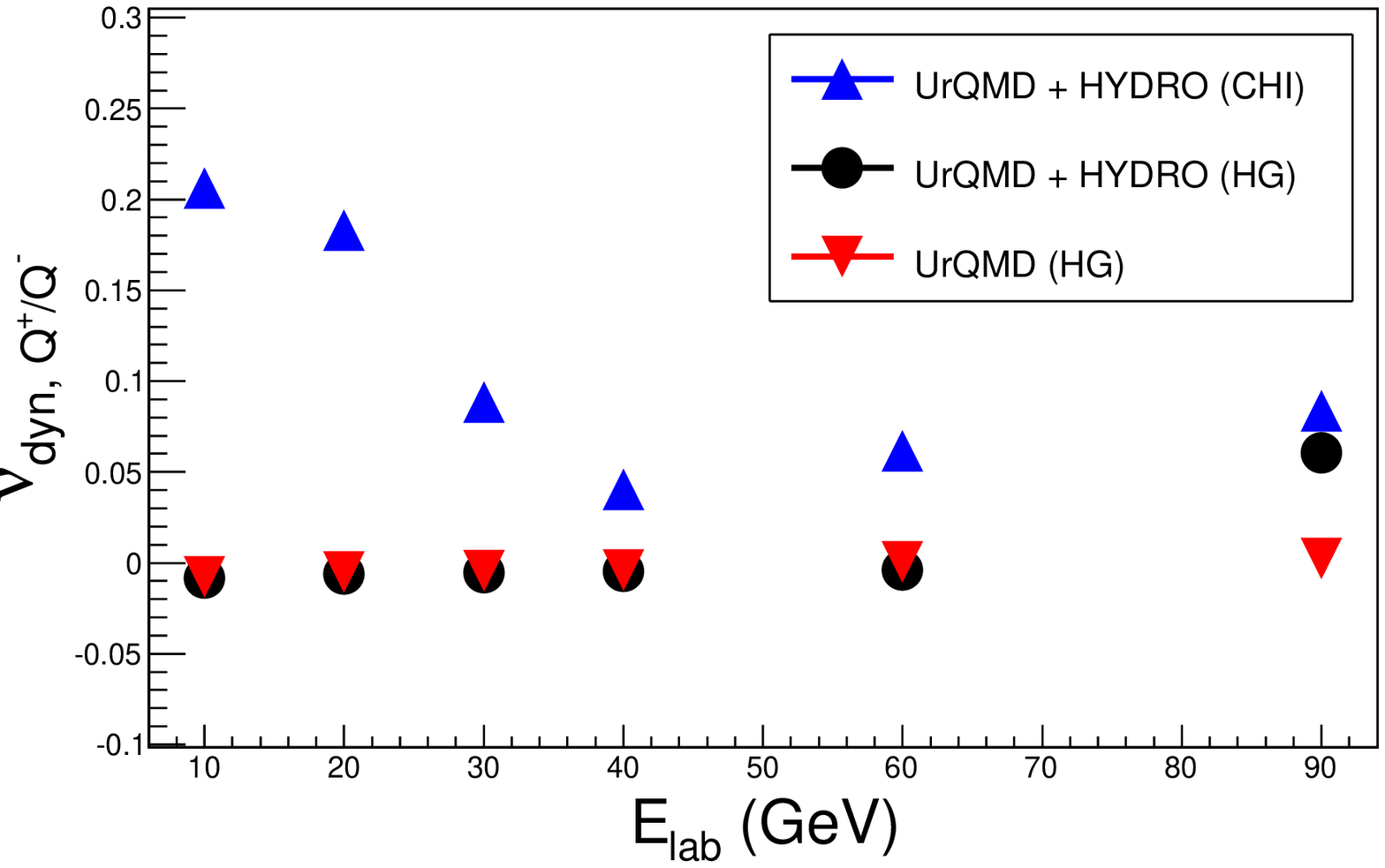} 
  \caption{Beam energy dependence of $\nu_{dyn}$ for $K/\pi$ and Q$^+$/Q$^-$ with pure transport and with hydro evolution using two equation of states}
  \label{hyd-two-eos}
  \end{figure}

In Fig.~\ref{hyd-two-eos} we have shown the energy dependence of $\nu_{dyn}$ at freezeout for three scenarios i.e., pure hadronic transport case with no hydrodynamic expansion, with hydrodynamic expansion using "hadronic gas" equation of state and then with "chiral" equation of state. It is seen that hydro expansion with hadronic equation of state is similar to the case of pure transport results. Chiral equation of state however, shows a different energy dependence for both the cases of fluctuations i.e., K/$\pi$ and Q$^+$/Q$^-$ showing clear enhancement at lower beam energy. This makes the measurement of fluctuation at lower beam energy more interesting.

\section{Fluctuations of net-proton distributions}
As mentioned earlier, measurements of higher order moments of the conserved quantities have drawn a lot of attention recently due to their higher sensitivity to the exotic phenomena like critical point. As a candidate of net-baryon numbers, results have been presented for net-proton distributions by several experiments. We have therefore extracted the time evolution of the net-proton and its fluctuation measured in terms of two widely used observables which represent the susceptibilities i.e., K$\sigma^2$ and S$\sigma$, where K, S and $\sigma^2$ represent kurtosis, skewness and variance of the distributions. Here we have taken the same coverage  and kinematic cuts as that of STAR published results~\cite{sk3}, i.e. psudorapidity of 0.5 around mid-rapdity and $p_{T}$ in the range of 0.4 - 0.8 (GeV/c), for the comparison of the results at freezout. It has been shown that these combinations of variables could be compared directly with the lattice simulation results.

\begin{figure}[h]
  \includegraphics[width=0.45\textwidth]{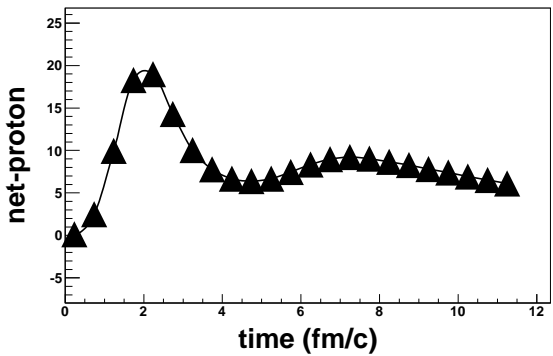} 
  \caption{Time evolution of the mean of net-proton at E$_{lab}$ = 40 AGeV. Like other observables, it shows a peak at about 2 fm/c which also depicts the peak of the net-baryon density. }
  \label{netp-time}
  \end{figure}
 
Fig.~\ref{netp-time} represents the time evolution of net-proton at E$_{lab}$ = 40 AGeV. As expected, at lower collision energy, baryons are stopped in the collision zone and the stopping is higher at the time of overlap of the colliding nuclei. As the time progresses, particles are produced and the net-proton density reduces to a saturated value. It is clearly seen that net proton shows a peak at around 2 fm/c, the time of duration of overlap.

\begin{figure}[h]
  \includegraphics[width=0.45\textwidth]{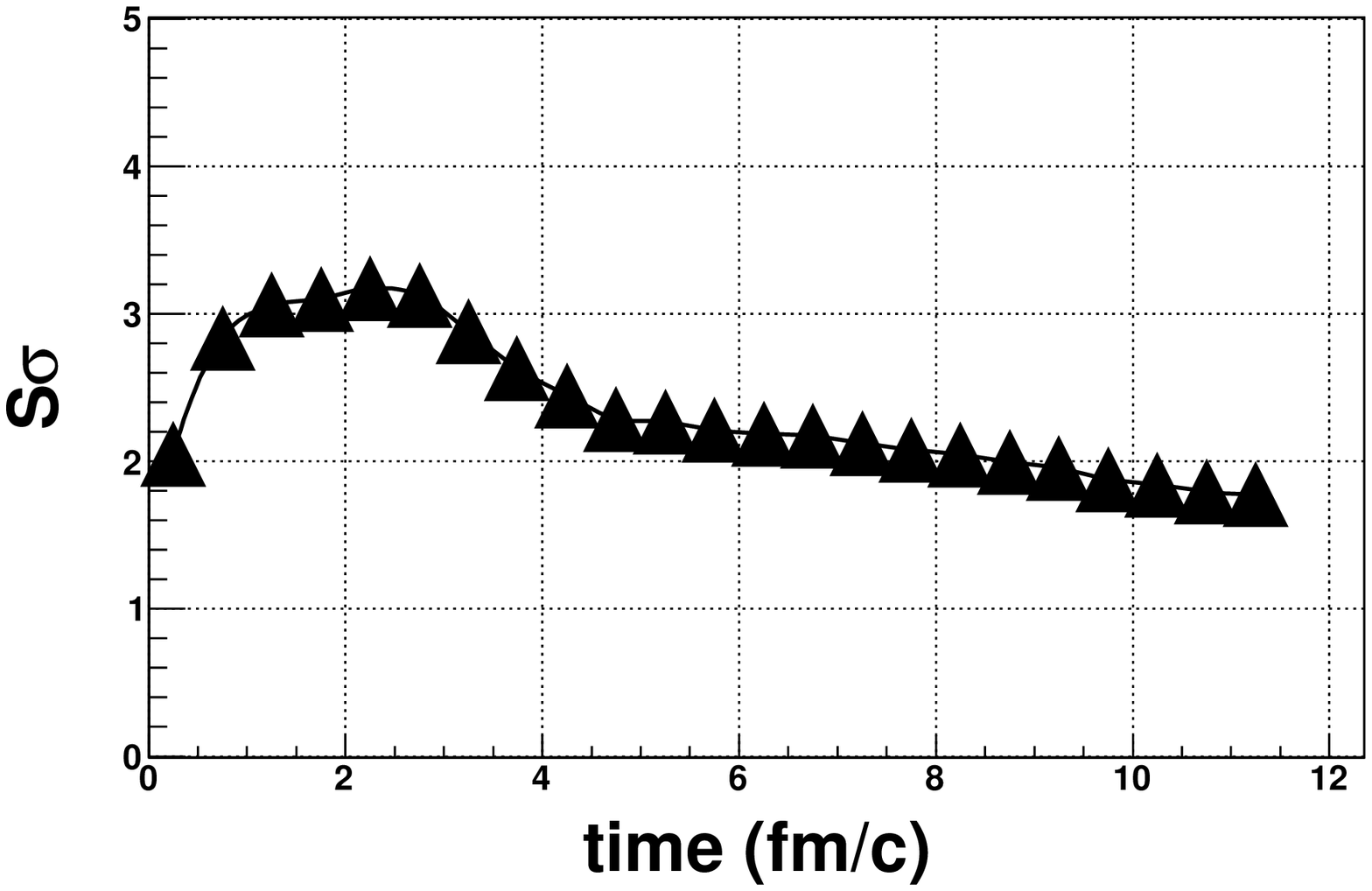} 
\includegraphics[width=0.45\textwidth]{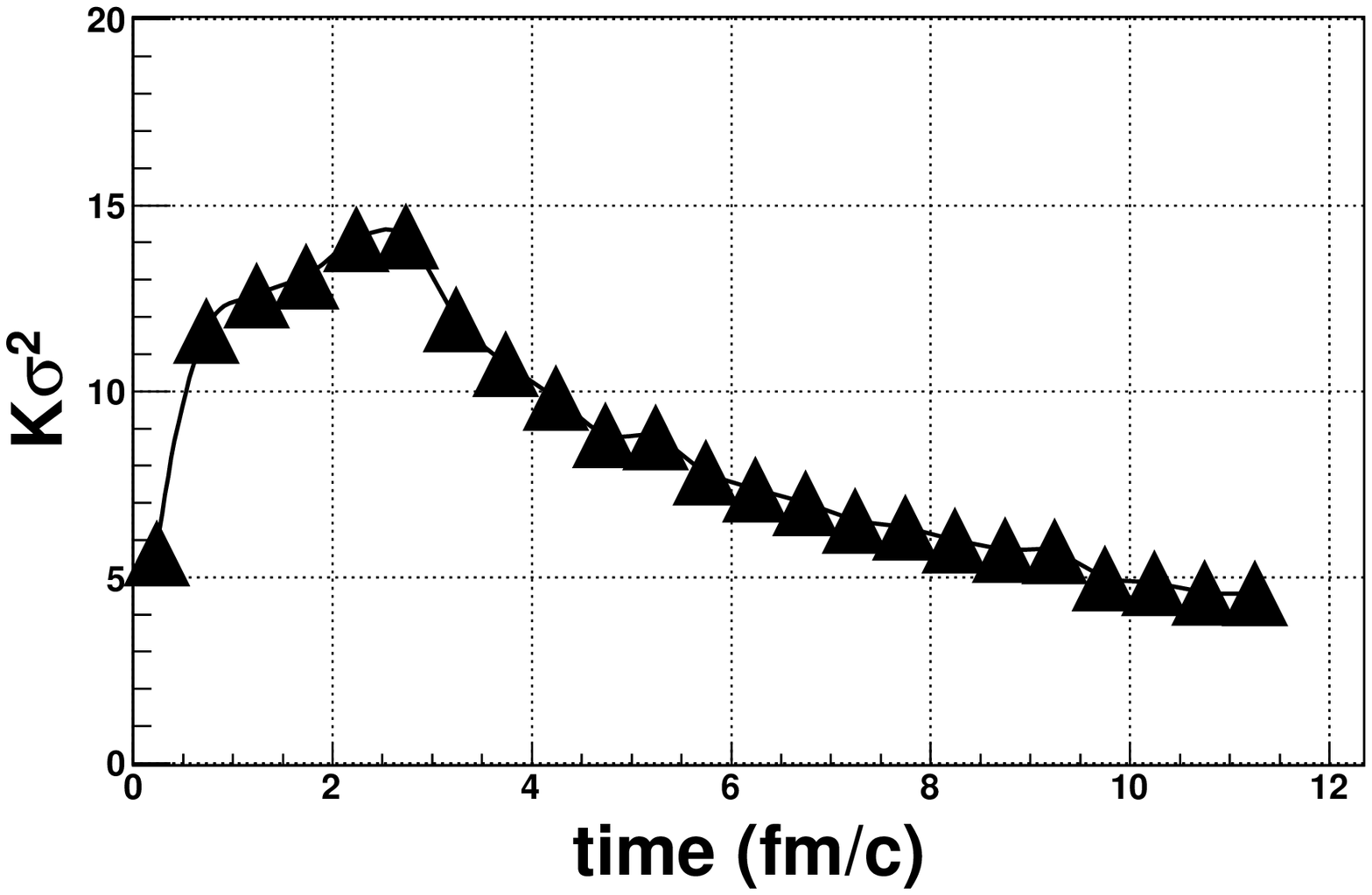} 
  \caption{Time evolution of S$\sigma$ (top) and K$\sigma^2$ (bottom) for net-proton at E$_{lab}$ = 40 AGeV.}
  \label{higher-mom-time}
  \end{figure}
 
\begin{figure}[h]
  \includegraphics[width=0.45\textwidth]{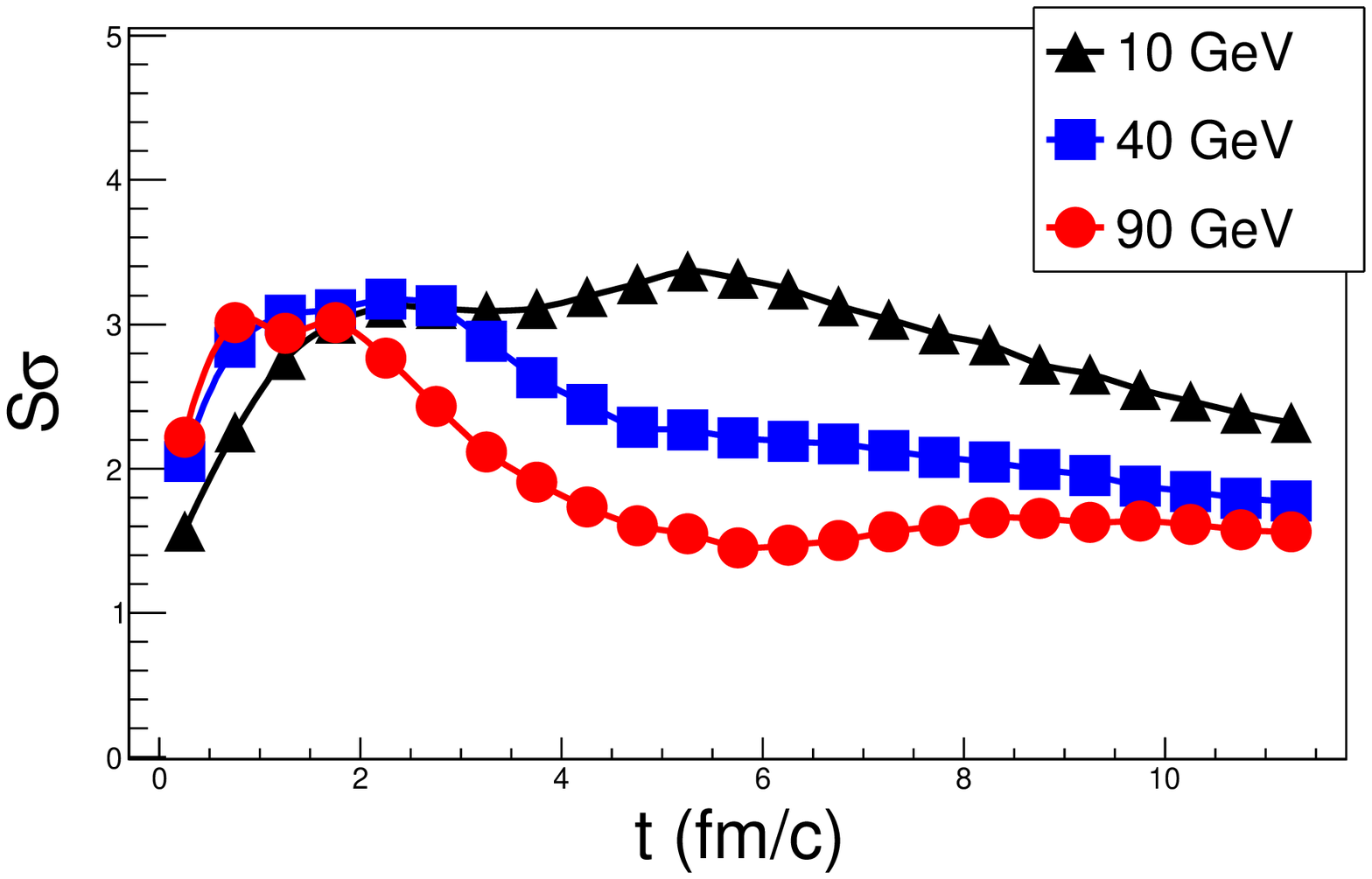} 
\includegraphics[width=0.45\textwidth]{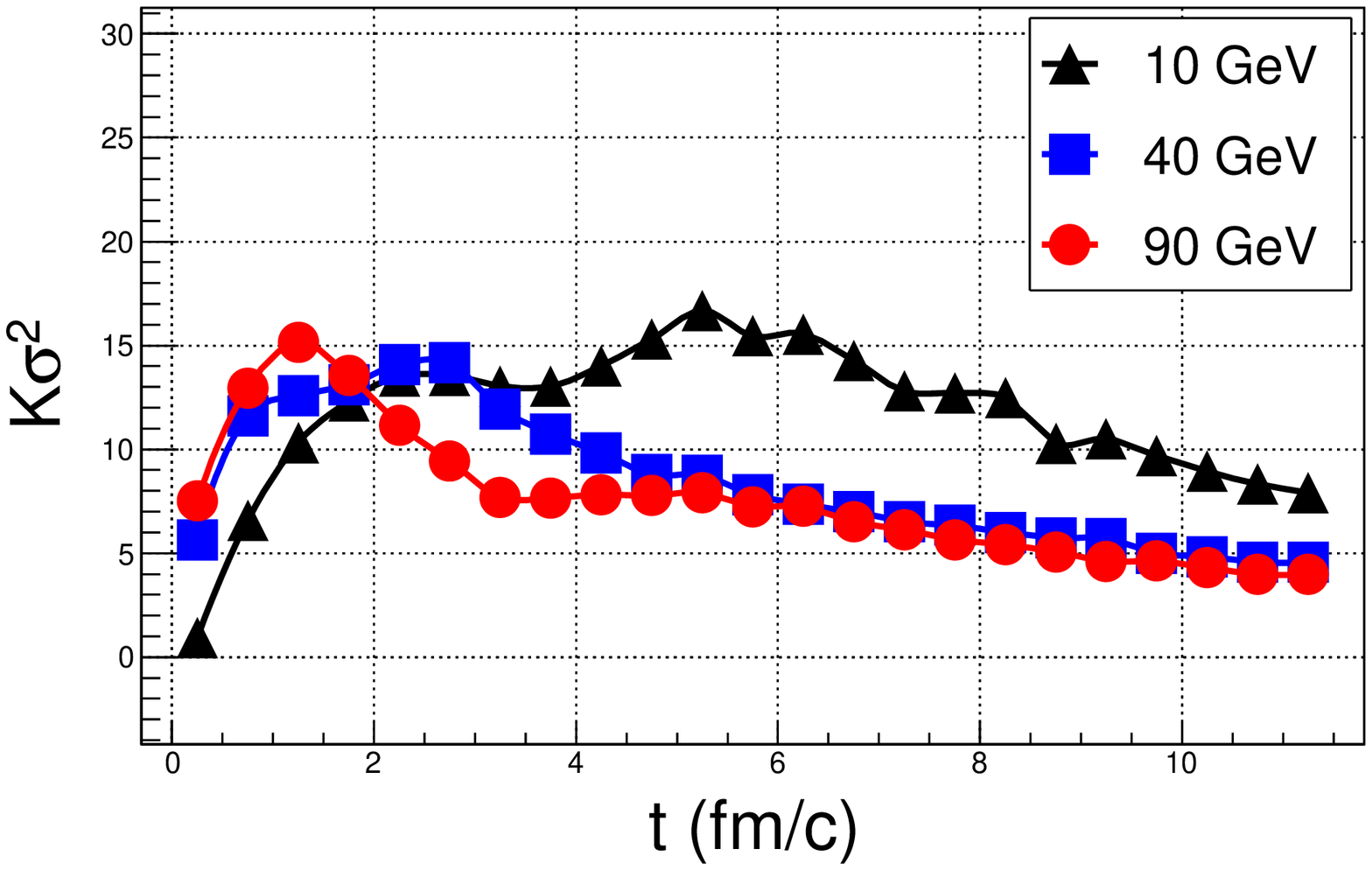} 
  \caption{Time evolution of S$\sigma$ (top) and K$\sigma^2$ (bottom) for net-proton at different collision energies}
  \label{higher-mom-elab}
  \end{figure}

Fig.~\ref{higher-mom-time} shows the time evolution of S$\sigma$ and K$\sigma^2$, two widely used variables consiting of the combinations of the higher order moments for net proton. It is seen that with higher baryon stopping, the fluctuations also increase and then reduces to saturated values towards freezeout. Like other fluctuation measures, these two observables as measred at freezeout is a result of significant dilution. In the study of fluctuation, we look for break from monotonous beahaviour of the fluctuation measures. In recent beam energy scan results from STAR, $\sqrt{s}$ dependence  of these quantities have been shown and some intesting structure at around 19 GeV have been mentioned about. It is evident from the present study that the values measured at freezeout does not contain the structures which might have been present during the evolution. We therefore would like to see the fate of the energy dependence of the structures. Towards this, we have plotted the beam time evolution at different beam energies as shown in Fig.~\ref{higher-mom-elab}. It also shows that at lower energy, the values are higher for the variables concerned but the trend is converging.

\begin{figure}[h]
  \includegraphics[width=0.45\textwidth]{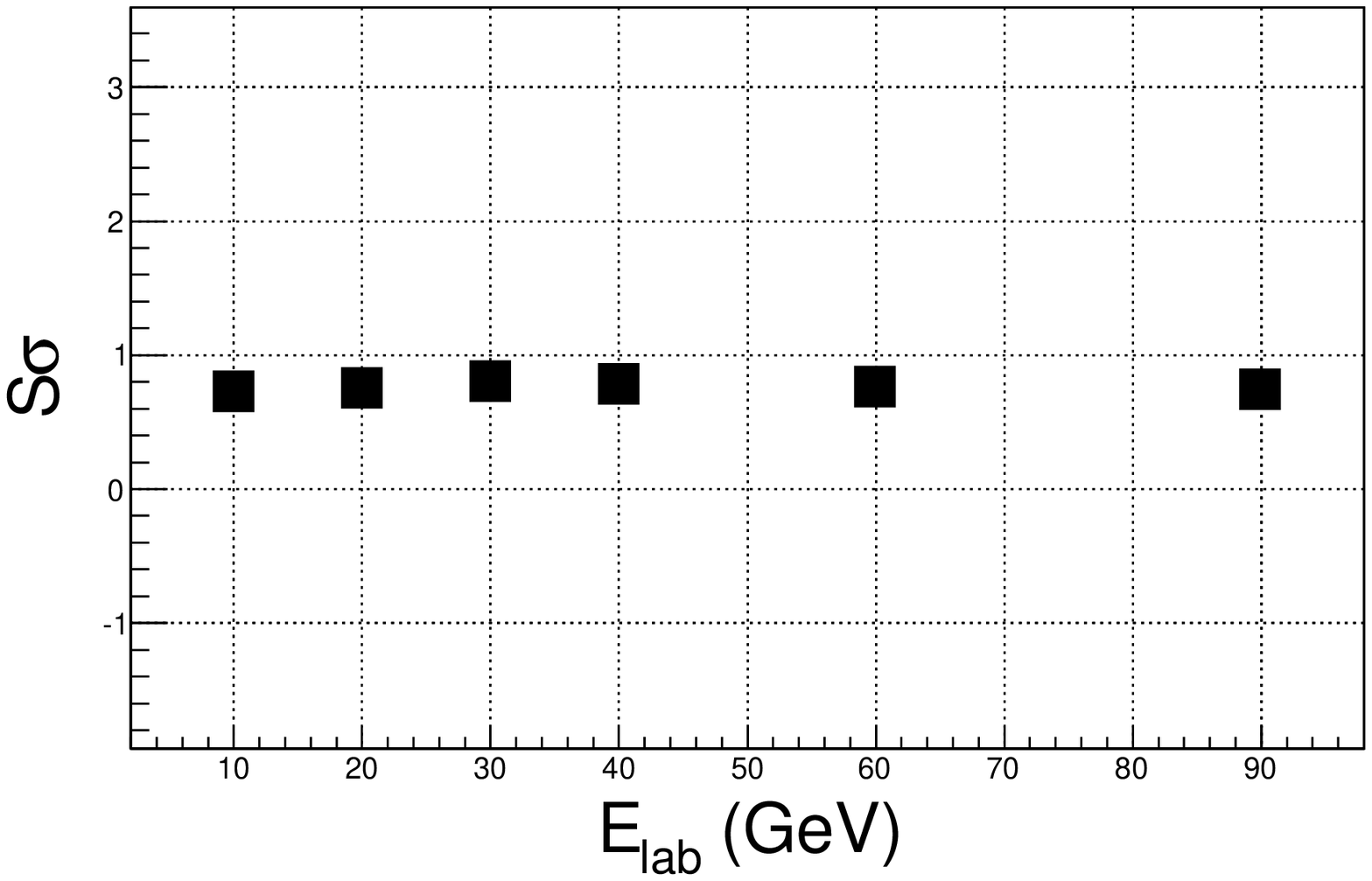} 
\includegraphics[width=0.45\textwidth]{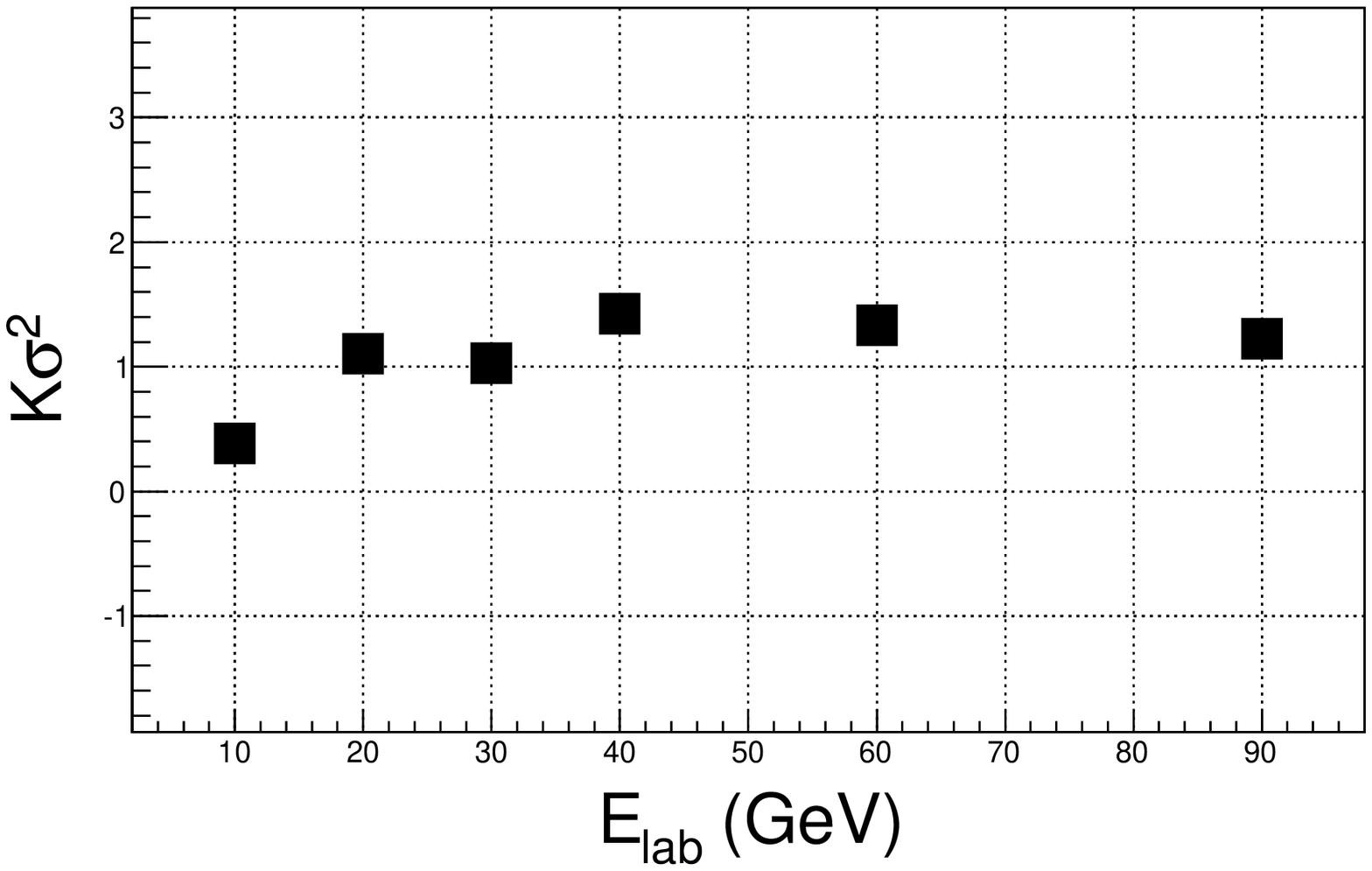} 
  \caption{Beam energy dependence of S$\sigma$ (top) and K$\sigma^2$ (bottom) for net-proton at freezout}
  \label{higher-mom-elab-freez}
  \end{figure}

Fig.~\ref{higher-mom-elab-freez} shows the energy dependence of S$\sigma$ and K$\sigma^2$ at freezout. Results are found to be consistent with the published results of STAR experiment~\cite{sk3}, there is almost no non-monotonius depence with $E_{lab}$ even at lower FAIR-energies within the same coverage. This is on an expected line as UrQMD does not have any physics which might create such an effect.

\section{Discusssions}
In this work, we have used UrQMD model to study the fluctuations of two important variables i.e the ratios of  K/$\pi$ and Q$^+$/Q$^-$ and the higher order moments of net-proton. The fluctuations of these variables are being studied extensively in high energy heavy-ion collisions as they are to signify important phenomena like phase transition and occurence of critical point. The results from UrQMD and other transport models have been used in the context of explaining the data from different experiments. The results from these models have been considered as reference to the data. All these results have been obtained at freezeout for comparison with data in equal footing. The transport models however have the advantages of providing space time history of the collision and can be used to understand the time evolution of signals till freezeout. The fluctuations are usually discussed in literature only in the context of thermal production of particles, however, if a vaiable shows very large fluctuation in the pre-equilibrium stage, then it needs to be understood at freezeout. 
In the present study, we have put emphasis on lower energy heavy-ion collisions likely to be availble at FAIR. These collisions produce matter under very high net-baryon density. It is seen from the transport models that the net baryon density reaches maximum for a short time during the overlap of two nuclei. During this time, the particle multiplicity also reaches  its peak and then slowly through the evolution reaches a near-saturated values at freezeout. During this evolution, flutuations of both the multiplicities and their ratios show structures consisting of peaks and valleys which then smoothen out during the evolution. It therefore suggests that there is considerable modification of the early fluctuation signals. From this analysis however, we might not be able to infer about the fluctuation signals which might have originated in an equilibrated thermal system. We have studied the effect of hydrodynamical evolution on those two signals as studied by evoluation of the signals at freezeout with transport only, hydro with hadronic EOS and chiral EOS respectively. It is seen that at freezeout, $\nu_{dyn}$, the measures of fluctuations of K/$\pi$ and Q$^+$/Q$^-$ ratios for transport only and hydro with hadronic EOS show similar results, however, the fluctuations with the chiral EOS shows a strructure in which the $\nu_{dyn}$ shows much larger values at lower beam energy. This therefore suggests that even at freeze-out, these might be considered to be an important signal for a specific EOS. For net-proton, fluctuation observables like S$\sigma$ and K$\sigma^2$ shows the similar trend at lower energies as published by the STAR at higher energies. But evolution of S$\sigma$ and K$\sigma^2$ shows initial fluctuation, at the time of nuclear overlap, which washes out with time with a converging trend.


\end{document}